\pdfoutput=1 

\documentclass[twocolumn,showpacs, 
amsmath,amssymb]{revtex4-1}

\usepackage{graphicx}
\usepackage{bm}

\input amssym.def
\input amssym
\newcommand{\be}{\begin{equation}}
\newcommand{\ee}{\end{equation}}

\begin{document}

\title{On a new category of physical effects}
\author{Nikolai V. Mitskievich}
\surname{Mitskievich}

 \email{nmitskie@gmail.com}

 \affiliation{Physics Department, CUCEI,
University of Guadalajara, Guadalajara, Jal., Mexico.}

\date{March 2010}
\begin{abstract}

A new category of ``intrinsic'' effects is proposed to be added to
the two already known kinematic and dynamical categories. An
example of intrinsic effect is predicted, its origin source is
established, and a scheme of its experimental detection is
proposed. This effect lowers to non-relativistic values the
propagation velocity of a plane electromagnetic wave in a vacuum,
when a time-independent homogeneous magnetic field is superposed
over it. This result, pertaining to the classical Maxwell theory,
follows from exact calculations. A critical remark on
gravitational waves' detection is given.
\end{abstract}



\maketitle

``In 1889 Morley and Miller... reconfigured the old
Michelson--Morley apparatus to search for changes in the speed of
polarized light caused by a magnetic field.''--- With this
historic reference begins the paper by Frank Nezrick \cite{Nezr}.
Morley and Miller, as well as Nezrick and his group at the
Fermilab, have considered dynamical mechanisms in their
predictions, in the Nezrick case, of the modern quantum field
theoretical origin, and the author categorically summarizes that
``[the] unquantized Maxwell Equations are linear in the fields
giving no interaction between photons.''

Below we explicitly show that the linearity of classical Maxwell's
equations in vacuum does not hinder from the existence of an
interaction between superposed electromagnetic fields in spite of
their perfectly exact and simple superposable nature. It is well
known that certain properties of these fields are determined by
the non-linear electromagnetic energy-momentum tensor $T^\mu_\nu$,
further taken in the classical vacuum. Our scientific community
has a deep-seated tradition to misapprehend as a nonsense the
natural situation when the same theory states that a superposition
of two or more exact solutions is itself an exact solution, {\it
and} it meanwhile predicts an intrinsic effect which {\it does not
inflict changes} in the component parts of this superposition,
while its physical characteristics cannot be reduced to a sum for
these component parts. This new classical intrinsic effect follows
from such non-linear and bilinear things as the energy density and
the Poynting vector whose combination, even {\it more} non-linear,
was already related by several authors to the observable group
velocity of electromagnetic field's propagation. They were Pauli
\cite{Pauli}, p. 115, Eq. (312): $\mathbf{v}^i=2T^i_0/T^0_0$;
Landau and Lifshitz \cite{LanLif}, the Problem on p. 69:
$\mathbf{v}^i/(1+v^2)=T^i_0/T^0_0$, dealing (from the authors'
viewpoint) only with parallelization of the electric and magnetic
fields; Penrose and Rindler \cite{PenRin}, vol. 1, p. 324, and
vol. 2, pp. 33 and 257-258, who considered only the pure electric
and magnetic type fields ({\it cf.} \cite{Mitsk08}), eliminating
the alternative 3-field (transformation to single-field frames).
The corresponding boosts are $B^{-2}(\mathbf{E}\times \mathbf{B})$
and $E^{-2}(\mathbf{E}\times \mathbf{B})$, respectively. Thus
Penrose--Rindler's boosts look as mixtures of the Poynting vector,
taken in the non-co-moving frame, but divided by energy density of
the electromagnetic field, pertaining to the frame co-moving with
this field. The velocity we are speaking here about is in fact
related to that which had to be measured by Frank Nezrick, though
he didn't mention the above authors. All this now occurs in the
non-quantized electromagnetic theory, thus we do not bother about
an ``interaction between photons.'' This situation has to reappear
also in general relativity and in quantum mechanics (see below),
but in special relativity the same type of effect has to be
present, easily calculated, and immediately detectable: see below
our computation.

Working on this effect in the special-relativistic Minkowskian
space-time, we shall use the $(+,-,-,-)$ signature and natural
units (so that the velocity is dimensionless and the
velocity-of-light constant is $c=1$), the Gaussian units in
Maxwell's equations, Greek 4-dimensional indices, and the Cartan
formalism of exterior forms (see \cite{Isr}) as the simplest and
most effective way to treat geometric ideas and to interpret the
obtained results. The electromagnetic field tensor splits, with
the help of monad $\tau$ (a unitary time-like vector field, in
fact, the 4-velocities of local test observers) and the dual
conjugation (or its Hodge-star form), into two 4-dimensional
(co)vectors, electric and magnetic, both $\perp\tau$:
\begin{widetext}
\begin{align}
\label{BE} \mathbf{E}_\mu =F_{\mu\nu} \tau^\nu ~ ~ ~ ~
\Leftrightarrow ~ ~ \textnormal{{\bf E}}= \ast(\tau \wedge\ast F),
~ ~ ~ ~ \mathbf{B}_\mu=-F\!\!\stackrel{\textnormal{\small$\ast$}
}{\textnormal{\scriptsize$\mu\nu$}}\!\tau^\nu ~ ~
\Leftrightarrow ~ ~ \mathbf{B}=\ast &(\tau\wedge F),\\
\label{FBE}  F =\mathbf{E}\wedge\tau+ \ast(\mathbf{B}\wedge\tau),
~ \ast F = \ast(\mathbf{E}\wedge\tau)- &(\mathbf{B}\wedge\tau).
\end{align}
\end{widetext}
It is obvious that {\bf E} is a polar 4-covector and {\bf B}, an
axial 4-covector, both restricted to the local physical 3-subspace
of the $\tau$-reference frame. The deduction details of the above
formulae see next to Eqs. (3.1.13), (3.1.16), and (3.1.18) in
\cite{Mitsk06}. In that book, the complete monad theory of
physical reference frames is given. J{\"u}rgen Ehlers
\cite{Ehlers} was first who formulated the monad formalism; with
it he worked exclusively in the cosmology using reference frames
co-moving with matter. The monad belonging to such frames he
denoted as $u$ which coincides with the 4-velocity of the filler
of cosmological space, so that we use here this notation also for
a monad (if any) co-moving with the electromagnetic field. A
general monad we denote as $\tau$, mostly since the integral lines
of the vector field $\tau$ are the physical time (not necessarily
time coordinate) lines in a space-time diagram. The second author
who independently formulated the monad formalism was Abram L.
Zel'manov \cite{Zelm}, and he worked with it also in relativistic
cosmology, like Ehlers.

A combination of $T^\alpha_\beta$ with an arbitrary monad $\tau$
yields
\begin{widetext}
\be \label{Ttausq} T^\nu_\mu T^\mu_\xi\tau_\nu \tau^\xi
=\frac{1}{(8\pi)^2}\left[\left(\textnormal{{\bf E}}^2+
\textnormal{{\bf B}}^2\right)^2-4(\textnormal{{\bf E}}\times
\textnormal{{\bf B}})^2\right]
\equiv\frac{1}{(8\pi)^2}\left[\left(\textnormal{{\bf B}}^2-
\textnormal{{\bf E}}^2\right)^2+4(\textnormal{{\bf E}}\bullet
\textnormal{{\bf B}})^2\right]=\frac{1}{(16\pi)^2}\left({I_1}^2+
{I_2}^2\right) \ee
\end{widetext}
where $I_1=F_{\mu\nu}F^{\mu\nu}$ and
$I_2=F\!\!\stackrel{\textnormal{\small$\ast$}
}{\textnormal{\scriptsize$\mu\nu$}}\! F^{\mu\nu}$ are two
invariants on which the simplest classification of electromagnetic
fields (see \cite{Mitsk08}) is based. It is remarkable that these
constructions are not only scalars under general transformations
of coordinates, but they are also independent of the reference
frame choice: the right-hand side does not involve the monad.
Considering the propagation of electromagnetic field, we do not
include the high-frequency limits related to field's
discontinuities (bicharacteristics). From (\ref{Ttausq}) and the
Landau and Lifshitz 3-velocity taken as an example, we see that
\begin{widetext}
\be \label{modv} 0\leq\frac{|\mathbf{v}|}{1+\mathbf{v}^2}=
\frac{1}{2} \sqrt{1-\frac{{I_1}^2+{I_2}^2}{4(\mathbf{E}^2+
\mathbf{B}^2)^2}}=\frac{|\mathbf{E}||\mathbf{B}|}{\mathbf{E}^2+
\mathbf{B}^2}|\sin\alpha|\leq\frac{1}{2}, \ee
\end{widetext}
$\alpha$ being the angle between {\bf E} and {\bf B} in the strict
local Euclidean sense; moreover, the function
$|\mathbf{v}|/(1+\mathbf{v}^2)$ is everywhere monotonic. In
particular, this means that the propagation of all pure null
fields ($|\mathbf{E}|=|\mathbf{B}|$, $\alpha=\pi/2$) occurs with
the unit absolute value of the 3-velocity, the velocity of light,
and all other electromagnetic fields propagate with sub-luminal
velocities which can always be made equal to zero in corresponding
co-moving reference frames. These conclusions also hold in general
relativity, and they are universally expressed in seemingly
``3-dimensional'' notations characteristic to the general
reference frame theory.

Instead of taking any of the 3-velocities given in \cite{Pauli,
LanLif, PenRin}, we shall now use our general definition
(\cite{Mitsk06}, p. 42, Eq. (2.2.11)) of the exact velocity
$\mathbf{v}$ between $u$-monad and $\tau$-monad frames from the
viewpoint of the latter frame: \be \label{v} u=(\tau\cdot u)(\tau+
\mathbf{v}), ~ ~ u\not\perp\mathbf{v}\perp\tau. \ee It is obvious
that this $\mathbf{v}$ will be the desired velocity of the
electromagnetic field propagation in $\tau$-frame, if the Poynting
vector vanishes in $u$-frame. Thus let us consider a linearly
polarized plane monochromatic electromagnetic wave (the situation
does not substantially depend on this choice of polarization) as
the first component of the superposition with a time-independent
homogeneous magnetic field (the second component) in the
$z$-direction of propagation of the wave (a constant electric
field yields similar results). In Cartesian coordinates $t,x,y,z$
(the spatial ones forming a right triplet), this superposition
reads (since $\mathbf{E}$ and $\mathbf{B}$ are covectors $\perp
dt$, the negative signs meaning positivity of the respective
vectors' components; to make all expressions more concise, we
abbreviate $\omega(t-z)$ as phase of the wave $\Theta$ and $E/H$
as $A$): \be \label{sup} \mathbf{E}=-E\cos\Theta dx, ~ ~
\mathbf{B}=-Hdz-E\cos\Theta dy \ee where $E$ is the scalar
amplitude of both electric and magnetic vectors of the wave, and
$H$ is the constant (both in space and time) magnitude of
superimposed magnetic field. For two electromagnetic invariants
one easily finds that $I_1:=2(\mathbf{B}^2-\mathbf{E}^2)=2H^2>0$,
$I_2:=-4\mathbf{E}\cdot\mathbf{B}=0$; consequently, this field
belongs to the pure magnetic type (see details of the
classification in \cite{Mitsk08}). This superposition is in fact a
specific not precisely monochromatic wave whose behavior can be
best understood in the reference frame co-moving with it. We shall
find such a frame using the pure-magnetic-type property of this
wave's field. First, we write the field $\ast F$ as a {\it simple}
bivector (see \cite{Isr}, p. 26). Taking for the general frame
monad $\tau=dt$, we find from (\ref{FBE}) and (\ref{sup}) that
\begin{align} \nonumber \ast F &=\left(E\cos\Theta dz \wedge
dy+H\phantom{^Z}\!\!\! dz\wedge dt+E\cos\Theta dy \wedge dt
\right)\\ &\equiv \underbrace{\left(Hdz+E \phantom{^Z}\!\!\!
\cos\Theta dy \right)}_{P,\textnormal{ a spacelike covector
}}\wedge \underbrace{(dt-dz)}_{Q,\textnormal{ a null covector
}}=P\wedge Q. \label{astFsup} \end{align}

\begin{widetext}\begin{center}\begin{figure}[h]
\hspace*{-14.4pt}\includegraphics[width=2.5in]{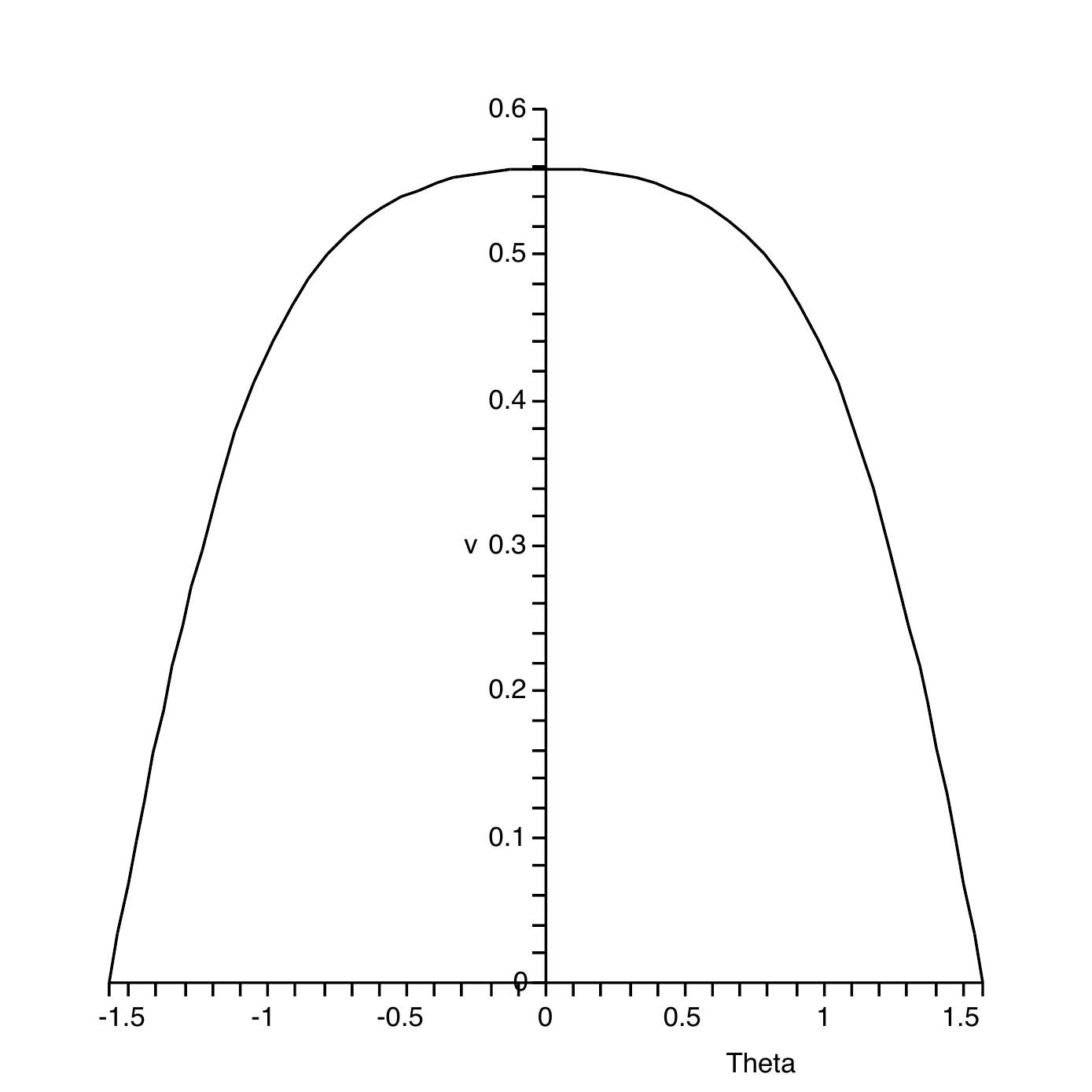}
\hspace*{-.25in}\includegraphics[width=2.5in]{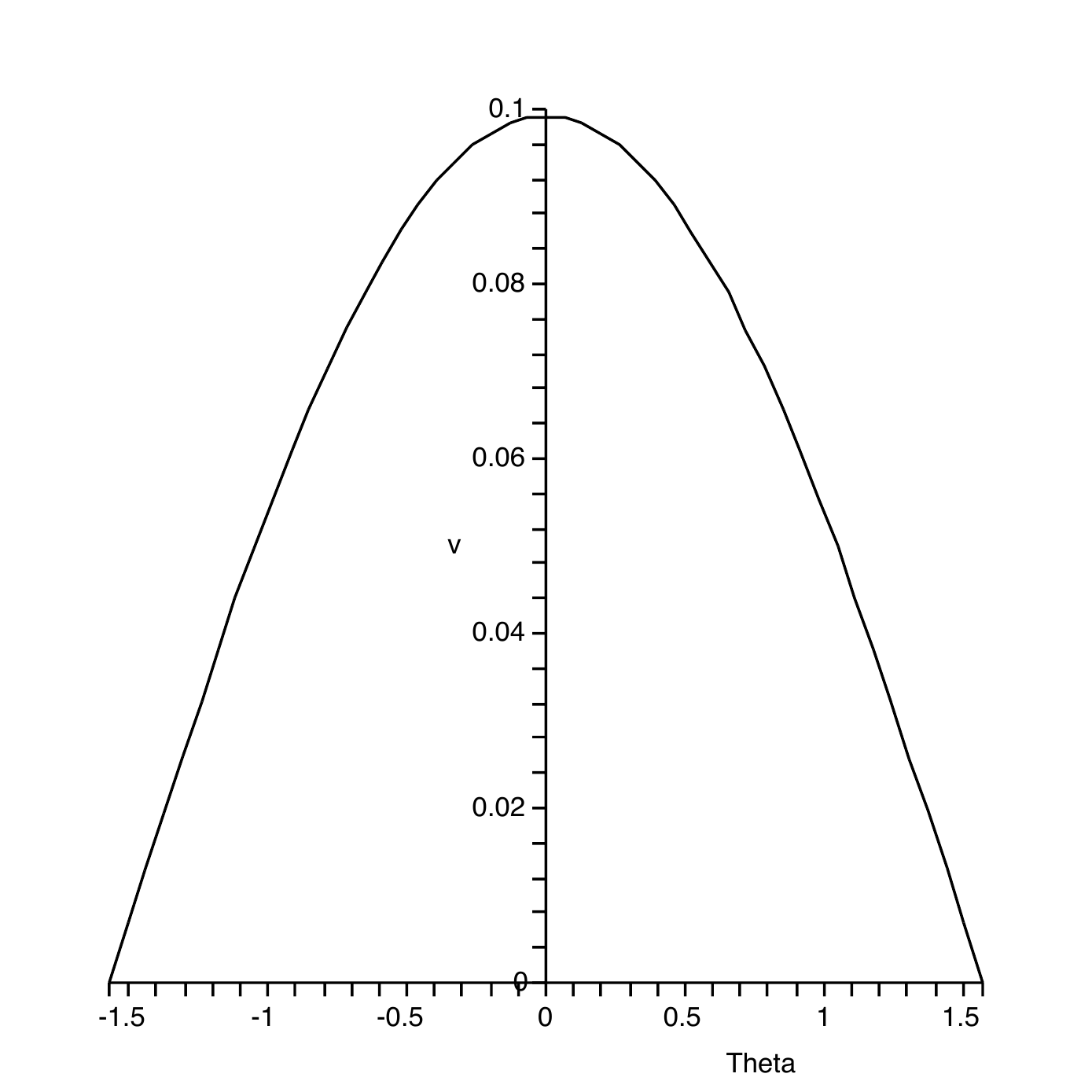}
\hspace*{-.25in}\includegraphics[width=2.5in]{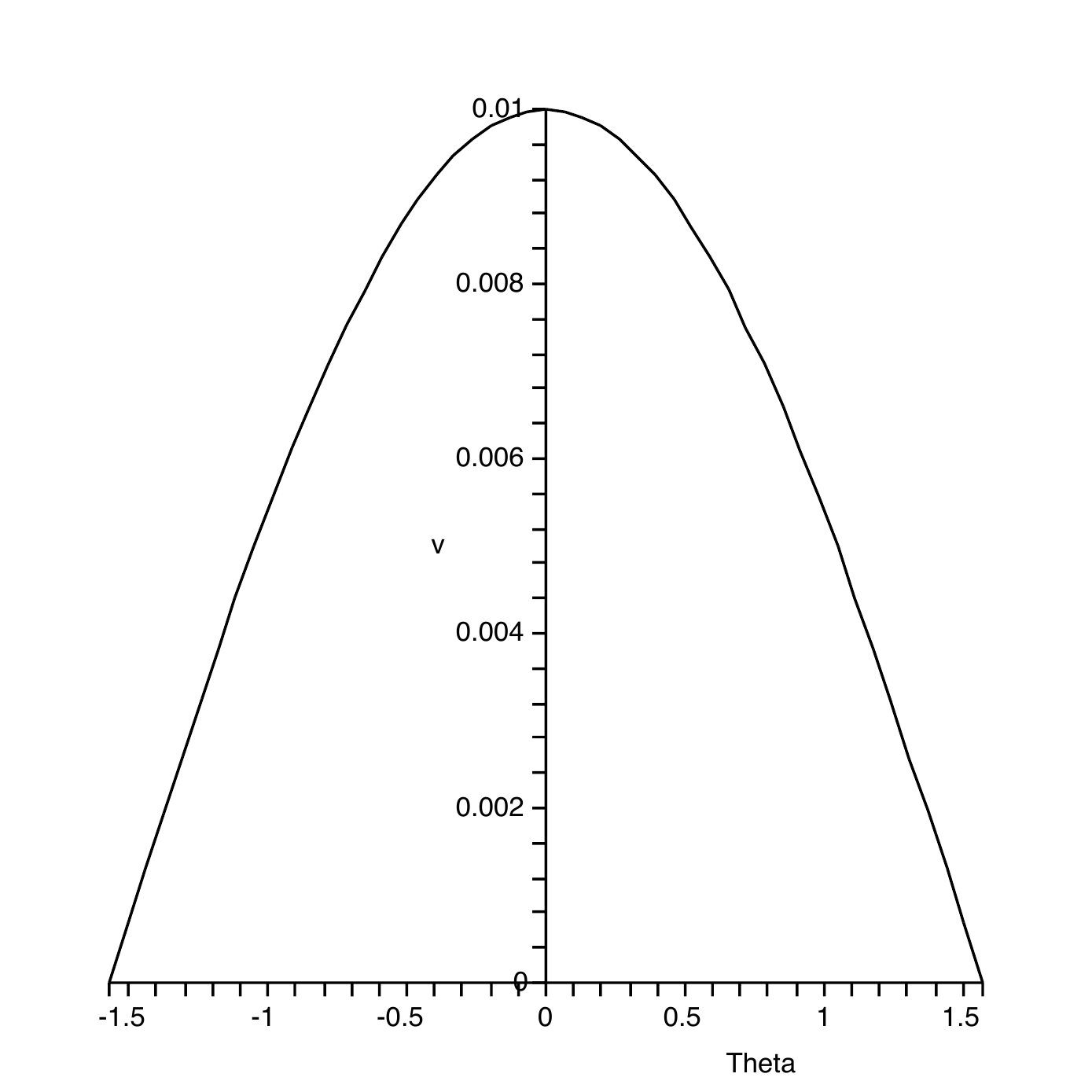}\\
\hspace*{-1.cm}{\phantom{aaa} (a) $E/H=1$ \hspace*{1.4in}(b)
$E/H=0.1$ \hspace*{1.4in} (c) $E/H=0.01$} \caption[Fig.
1]{Diagrams of the velocity $|\mathrm{v}|$ for
$-\pi/2\leq\Theta=$Theta$=\Theta\leq\pi/2$.}
\end{figure}\end{center}\end{widetext}

If to $P$ we add $lQ$ ($l$ being an arbitrary function) and use
this sum $P'$ instead of the former $P$, $\ast F$ does not suffer
any change (neither $F$ does). While $P\cdot P<0$, $P'\cdot
P'=2Hl-H^2-E^2\cos^2\Theta$. Thus if we choose
$l=H+\frac{E^2}{2H}\cos^2\Theta$, the vector $P'$ becomes
timelike, $P'\cdot P'=H^2>0$, and we can take $P'/H$ as a properly
normalized monad $u=\left(1+\frac{A^2}{2} \cos^2\Theta
\right)(dt-dz)+dz+A\cos\Theta dy$. Now (\ref{astFsup}) reads $\ast
F=Hu\wedge(dt-dz)$, so that in the new frame $u$ the electric
field identically vanishes due to (\ref{BE}) rewritten for $u$,
hence we have found one of the field's co-moving frames. In such
calculations one has to remember that when only one (here,
magnetic) field survives after the reference frame is transformed,
there are other possible transformations which already do not
change this situation (in fact, all those which involve an
additional motion in the direction of this 3-field, even when this
motion occurs to be with a non-constant magnitude of the
3-velocity described by strictly local Lorentz transformations,
thus working in non-inertial frames). Consequently, there appears
a continuum of such single-field ($\mathbf{E}$- or $\mathbf{B}$-)
frames ({\it cf.} \cite{PenRin}; of course, they work in general
as well as in the special relativity), and the search for more
elegant frames depends on the individual taste of the researcher.
This means that the 3-velocities given in \cite{Pauli, LanLif,
PenRin} may describe only particular choices of co-moving frames
(if they are correct at all).

Let us now calculate 3-velocity $\mathbf{v}$ of the co-moving
frame $u$ [group propagation velocity of the electromagnetic field
(\ref{sup})] with respect to the frame $\tau=dt$, using our
general definition (\ref{v}):
\begin{align} \label{velrel} \mathbf{v} &=\frac{A\cos\Theta}{1+
\frac{A^2}{2}\cos^2\Theta}\left(dy-\frac{A}{2}\cos\Theta dz\right),\\
\label{magnvel} |\mathbf{v}| &=
\frac{A\cos\Theta\sqrt{1+\frac{A^2}{4}\cos^2\Theta}}{1+\frac{A^2}{2}
\cos^2\Theta},
\end{align}
thus when $E/H\equiv A\leq 1$, $|\mathbf{v}|<1$, and frames
co-moving with the superposition of fields are, in principle,
realizable. When $H\rightarrow 0$, the propagation velocity
approaches to that of light, while if $E\ll H$, it becomes as low
as one wishes; see Fig. 1 where $|\mathbf{v}|$ is given in natural
units.

The most simply realizable experiment for detection of this
intrinsic effect can be performed with a large distance of the
light propagation in a sufficiently long optic fiber wound upon a
bobbin containing a straight concentric conductor with a direct
electric current in it to produce a magnetic field along the fiber
with the light beam. The dielectric properties of the fiber can be
easily filtered out since in this experiment the superimposed
constant magnetic field will dominate, and the comparatively weak
light beam can be manipulated to get different velocities of its
propagation, practically not changing the dielectric properties of
the fiber.

Of course, the effects expressed in mutually analogous
characteristics ({\it e.g}, those which are related to a shift of
the propagation velocity of electromagnetic field), can be
superimposed. However, in so doing, such parallel effects are of
entirely different orders of magnitude: the dynamical effects
depend on interaction constants (in particular, when there appears
a non-linearity in the dynamical field equations, for example, via
quantum theoretical corrections like those mentioned in
\cite{Nezr}); the kinematic effects ({\it e.g.}, the 3-velocities
composition law) are more universal, but they do not change $c=1$
composed with any subluminal velocity; the intrinsic effects are
significantly stronger than the other ones in view of the stable
non-linearity of the expressions which yield them, without any
participation of interaction constants. Therefore the intrinsic
effects generally are dominant also in other branches of physics.
The intrinsic effect related to the group velocity shift has to
occur in quantum mechanics (without the second quantization, in
the linear equations such as the Schr{\"o}dinger and Dirac ones)
where the 3-velocity should follow from the non-linear probability
density flow and probability density itself.

The gravitational deformation in general relativity does in fact
belong to the kinematic effects, when it is described without the
use of geodesic deviation equation (which would bring us back to
the weak dynamical Weber-type effect \cite{Weber}). Thus, for
example, the interferometric detection of gravitational waves
cannot give a non-zero result, since the scales of all types of
equally oriented lengths do change in gravitational fields in the
same proportion, and the numbers of light wavelengths fitting
along the alternative arms of interferometer cannot suffer changes
in a passing gravitational wave. I am regretful not to tell these
considerations to Kip S. Thorne more than two decades ago, simply
because of a kind of awkward modesty (at a seminar of the
Institute for Physical Problems in Moscow in 1970ies during the
talk of Herzenstein and Pustovoyt on their proposal of such a
detection of gravitational waves, when I had told them this fact,
the talk immediately collapsed, and I felt sorry for it).

I would like to thank Dr. Arturo Ch\'avez Ch\'avez for fruitful
discussions in which he acquainted me, in particular, with the
reference \cite{Nezr}.

\end{document}